\documentclass[twocolumn,showpacs,prb]{revtex4}
\usepackage{graphicx}
\usepackage{dcolumn}
\usepackage{bm}
\usepackage[section]{placeins}
\usepackage{float}

\begin{document}

\author{Z. Z. Zhang}
\affiliation{SKLSM, Institute of Semiconductors, Chinese Academy of Sciences, P.O. Box
912, Beijing 100083, China}
\author{Z. H. Wu}
\affiliation{SKLSM, Institute of Semiconductors, Chinese Academy of Sciences, P.O. Box
912, Beijing 100083, China}
\author{Kai Chang}
\email{kchang@red.semi.ac.cn}
\affiliation{SKLSM, Institute of Semiconductors, Chinese Academy of Sciences, P.O. Box
912, Beijing 100083, China}
\author{F. M. Peeters}
\affiliation{Department of Physics, University of Antewerp, Groenenborgerlaan 171, B-2020
Antwerpen, Belgium}
\date{\today }
\title{Resonant Tunneling through S- and U-shaped Graphene Nanoribbons }

\begin{abstract}
We theoretically investigate resonant tunneling through S- and U-shaped
nanostructured graphene nanoribbons. A rich structure of resonant tunneling
peaks are found eminating from different quasi-bound states in the middle
region. The tunneling current can be turned on and off by varying the Fermi
energy. Tunability of resonant tunneling is realized by changing the width
of the left and/or right leads and without the use of any external gates.
\end{abstract}

\maketitle

\section{INTRODUCTION}

Graphene is a single layer of carbon atoms arranged in a hexagonal lattice
that is the building block for graphite material. Recently, graphene samples
have been fabricated experimentally by micro-mechanical cleavage of graphite
\cite{Novoselov}. This material has aroused increasing attention due to its
novel transport properties that arises from its unique band structure: the
conduction and valence bands meet conically at the two nonequivalent Dirac
points, called $K$ and $K^{^{\prime }}$ valleys of the Brillouin zone, which
show opposite chirality. Around the two Dirac points, the energy dispersion
is linear in momentum space and is well described by the massless Dirac-Weyl
equation. The unique energy dispersion leads to a number of unusual
electronic transport properties in graphene such as integer quantum Hall
effect at room temperature \cite{QHE}, finite minimal conductivity, and
special Andreev reflection \cite{Beenakker1,Sengupta}. In graphene electrons
can pass through potential barriers without any reflection, i.e., the Klein
tunneling, in contrast to conventional electrons that exhibit an exponential
decreasing transmission with increasing barrier height and/or width. Klein
tunneling makes it impossible to confine the carriers in single layer\
graphene using the electric gates that are often used in the conventional
semiconductor two-dimensional electron gas (2DEG). Tunneling through single
barriers \cite{Katsnelson} and double barriers \cite{Pereira} in graphene
has been investigated theoretically. Recent theoretical work demonstrated
that the magnetic barriers can be used to confine the massless Dirac fermion
\cite{Egger} and  block the Klein tunneling and display a huge
magnetoresistance \cite{Zhai,Peeters}. Klein tunneling leads to a poor
rectification effect and the absence of resonant tunneling for normal
incidence in graphene and limits the performance of graphene-based
electronic devices.

In graphene nanoribbons (GNRs), the presence of edges can change the energy
spectrum of the $\pi $-electrons dramatically. GNRs can be fabricated using
conventional lithography and etching techniques \cite{IBM,kim}. The
electronic properties of a GNR depend very sensitively on the sizes and
shape of the edges, i.e., zigzag- and armchair-edged GNR. The zigzag-edged
graphene nanoribbons (ZGNRs) and armchair-edged graphene nanoribbons (AGNRs)
exhibit very different band structures. The ZGNRs always have localized
states appearing at the edge near the Dirac point, and therefore exhibit
metallic-like behavior. The AGNRs show alternating metallic-like and
semiconductor-like features alternatively as the width of nanoribbons
increases \cite{Nakada}. Such features are very different from the
conventional semiconductor quantum wire and provide a unique way to tailor
the transport properties of GNRs \cite%
{Wakabayashi,Wakabayashi2,Peres,Cresti,Areshkin,Yan}. Recently it is found
that single electronic potential barrier can block the electron tunneling in
the ZGNRs and confine electrons in between double barriers \cite%
{Wakabayashi2}. The magnetic transport and spatial distribution of currents
was investigated for unipolar and bipolar ZGNRs \cite{Cresti}. These
previous works have demonstrate that the resonant tunneling can be realized
using electric and/or magnetic barrier. Here, we will address the question
whether it is possible to realize resonant tunneling using \textit{only} the
geometry of the graphene nanoribbons. We focus on the S- and U-shaped
structures consisting of zigzag leads and armchair nanoribbons in between
them, because the zigzag nanoribbon is always metallic, while the bandgap of
an armchair nanoribbon depends sensitively on the width of nanoribbon. This
combination of zigzag and armchair nanoribbon could lead to interesting
transport property\textit{.}

In this work, we theoretically investigate the transport property through
U-shaped nanoribbon structures sandwiched by two semi-infinite ZGNRs, i.e.,
the geometry of the graphene nanoribbon in the absence of any external
nanostructured electric gates.. We will demonstrate resonant tunneling
behavior through these graphene nanostructures whenever the AGNR in the
middle part is metal-like or semiconductor-like. In those structures, the
middle structure acts as a barrier between ZGNR leads, and an interesting
resonant tunneling behavior can be seen with varying lengths of the middle
part of the U-shaped ZGNR or AGNR.

This paper is organized as follows. In Sec. II, the theoretical model and
calculation method are given. In Sec. III, we present the numerical results
and our discussions. Finally we give our conclusion in Sec. IV.

\section{MODEL AND CALCULATION METHOD}

To describe the electron transport through the \textit{S-} or U-like
structure sandwiched by two semi-infinite ZGNR leads, the tight-binding
Hamiltonian is adopted
\begin{equation}
H=-t\sum\limits_{<i,j>}\left( c_{i}^{\dagger }c_{j}+H.c.\right) -t^{\prime
}\sum\limits_{<<i,j>>}\left( c_{i}^{\dagger }c_{j}+H.c.\right) ,
\end{equation}%
where $c_{i\left( j\right) }^{\dagger }\left( c_{i\left( j\right) }\right) $
creates (annihilates) an electron on site $R_{i\left( j\right) }$, $t$ is
the nearest neighbor ($<i,j>$) hopping energy ($t\approx 3.03$ eV), and $%
t^{\prime }$ is the next-nearest neighbour ($<<i,j>>$) hopping energy \cite%
{Reich}. The conductance of the system is evaluated using the Landauer-B\"{u}%
ttiker formula\cite{butiker},
\begin{equation}
G\left( E_{F}\right) =\frac{e^{2}}{\pi }\sum\limits_{\mu }T_{\mu
}\left( E_{F}\right) ,\ \ T_{\mu }\left( E_{F}\right)
=\sum\limits_{\nu }\left\vert t_{\mu ,\nu }\left( E_{F}\right)
\right\vert ^{2},  \label{eq1}
\end{equation}

\begin{figure}[tbp]
\includegraphics [bb=0 0 1260 500,width=\columnwidth]{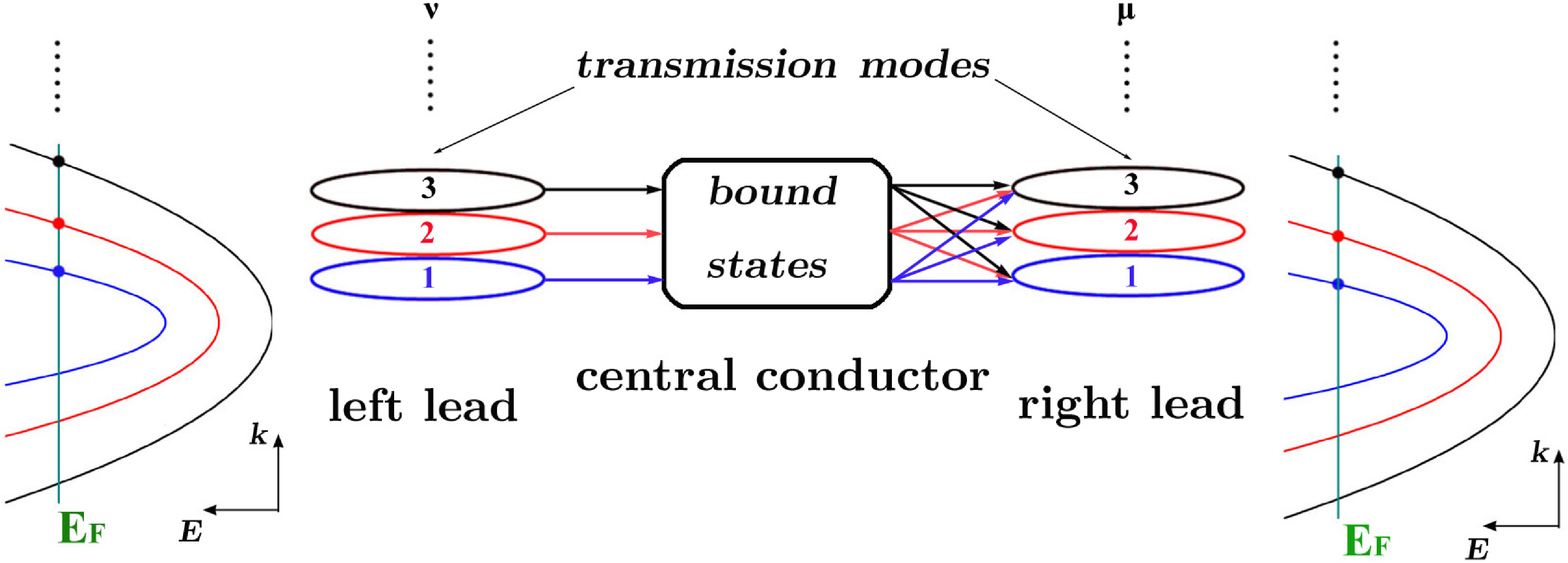}
\caption{(Color online ) Schematic diagram of the multi-moded transmission.}
\label{Fig7}
\end{figure}

where $t_{\mu ,\nu }\left( E_{F}\right) $ is the transmission coefficient
from the $\nu $-th channel in the left lead to the $\mu $-th channel in the
right lead at the Fermi energy $E_{F}$ (as shown in Fig.~\ref{Fig7}),
calculated by a recursive Green's function method \cite{Ando}, which is
given as
\begin{equation}
\boldsymbol{G_{S}}=\frac{1}{E-\boldsymbol{\tilde{H}}}=\frac{1}{E-\boldsymbol{%
H_{C}}-\boldsymbol{\Sigma _{L}}-\boldsymbol{\Sigma _{R}}},  \label{Greenfun}
\end{equation}%
where $E$\ is the Fermi energy, $\tilde{H}$\ is the effective Hamiltonian
that equals the sum of the Hamiltonian of the central rectangular ring, $%
H_{C}$, and the self energies of the semi-infinite leads in the left and
right leads, $\Sigma _{L}+\Sigma _{R}$. The key point of the recursion
Green's function method is that the middle conductor region is divided into
many meshes, and consequently reduce the dimensionality of the Green's
function greatly. Therefore the computation of the conductance becomes more
pratical.

The Fermi energy $E_{F}$ can be tuned experimentally through
electric top- and/or back-gates \cite{Ozyilmaz}. In our
calculations, all physical quantities are dimensionless, e.g., the
energy $E_{F}$ and the conductances are in units of $t$ and
$e^{2}/\left( \pi \right) $, respectively. The next-nearest
neighbour hopping term is ignored because it is much weaker than
that of the nearest neighbor hopping term, i.e., $t^{\prime
}=t/10^{4}$.

For completeness, here we repeat the essential steps in T. Ando's work \cite%
{Ando} to obtain recursive Green's function ~$G_{S}$~. The equation of
motion for an ideal lead can be written as:
\begin{equation}
(E-\boldsymbol{H}_{0})\boldsymbol{C}_{j}+\boldsymbol{TC}_{j-1}+\boldsymbol{T}%
^{T}\boldsymbol{C}_{j+1}=0,  \label{eq_motion}
\end{equation}
where $\boldsymbol{H}_{0}$ is the hamiltonian of a mesh in the ideal lead
containing M lattice sites, $\boldsymbol{C}_{j}$ is a vector describing the
amplitudes of the $jth$ mesh, $\boldsymbol{T}$ and $\boldsymbol{T}^{T}$ are
the matrix blocks adjacent to $\boldsymbol{H}_{0}$ in the total
tight-binding Hamiltonian matrix. To obtain linearly solutions for Eq.~\ref%
{eq_motion}, we first set $\boldsymbol{C}_{j}=\lambda ^{j}\boldsymbol{C}_{0}$%
. Substituting this in to Eq.~\ref{eq_motion} we have:
\begin{equation}
\lambda \boldsymbol{C}_{j}=\left( \mathbf{T}^{T}\right) ^{-1}\left(
\boldsymbol{H}_{0}-E\right) \boldsymbol{C}_{j}-\left( \boldsymbol{T}
^{T}\right) ^{-1}\boldsymbol{TC}_{j-1},
\end{equation}
leads to the following eigenvalue problem:
\begin{equation}
\lambda \left(
\begin{array}{c}
\boldsymbol{C}_{j} \\
\boldsymbol{C}_{j-1}%
\end{array}%
\right) =\left(
\begin{array}{cc}
\left( \boldsymbol{T}^{T}\right) ^{-1}\left( \boldsymbol{H}_{0}-E\right) &
-\left( \boldsymbol{T}^{T}\right) ^{-1}\boldsymbol{T} \\
\boldsymbol{1} & 0%
\end{array}%
\right) \left(
\begin{array}{c}
\boldsymbol{C}_{j} \\
\boldsymbol{C}_{j-1}%
\end{array}%
\right).
\end{equation}

We get the eigenvalues and corresponding eigenvectors, including M
right-going and M left-going waves. We define the matrix for the eigenvalues
and eigenvectors:
\begin{equation}
\boldsymbol{U}\left( \pm \right) =(\boldsymbol{u}_{1}(\pm ),\cdots ,%
\boldsymbol{u}_{M}(\pm ));
\end{equation}%
\begin{equation}
\boldsymbol{\Lambda }\left( \pm \right) =\left(
\begin{array}{ccc}
\lambda _{1}\left( \pm \right)  &  &  \\
& \ddots  &  \\
&  & \lambda _{M}\left( \pm \right)
\end{array}%
\right) .
\end{equation}%
Now, we get the relations between $j-th$ and $j^{^{\prime }}-th$ mesh:
\begin{equation}
\boldsymbol{C}_{j}\left( \pm \right) =\boldsymbol{F}(\pm )^{j-j^{\prime }}%
\boldsymbol{C}_{j^{\prime }}\left( \pm \right) ,  \label{eq_F}
\end{equation}%
with $\boldsymbol{F}(\pm )=\boldsymbol{U}\left( \pm \right) \boldsymbol{%
\Lambda }\left( \pm \right) \boldsymbol{U}^{-1}\left( \pm \right) $.

Next we consider the scattering in the middle conductor (from mesh 1to mash
N) to both sides of witch an ideal wire is attached. By using Eq.~\ref%
{eq_motion} and Eq.~\ref{eq_F}, we obtain the effective hamiltonians (note
that they are not hermitian) of the last mesh in left lead and first mesh in
right lead, both adjacent to the middle conductor:
\begin{eqnarray}
\boldsymbol{\tilde{H}}_{0}=\boldsymbol{H}_{0}-\boldsymbol{TF}^{-1}\left(
-\right) ,  \\
\boldsymbol{\tilde{H}}_{N+1}=\boldsymbol{H}_{0}-\boldsymbol{T}^{T}%
\boldsymbol{F}\left( +\right) .
\end{eqnarray}
We can get the total effective hamiltonian matrix:
\begin{equation}
\boldsymbol{\tilde{H}}=\left(
\begin{array}{ccccc}
\boldsymbol{\tilde{H}}_{0} & -\boldsymbol{T}^{T} & \ldots & \boldsymbol{0} &
\boldsymbol{0} \\
-\boldsymbol{T} & \boldsymbol{\tilde{H}}_{1} & \ldots & \boldsymbol{0} &
\boldsymbol{0} \\
\vdots & \vdots & \ddots & \vdots & \vdots \\
\boldsymbol{0} & \boldsymbol{0} & \ldots & \boldsymbol{\tilde{H}}_{N} & -%
\boldsymbol{T}^{T} \\
\boldsymbol{0} & \boldsymbol{0} & \ldots & -\boldsymbol{T} & \boldsymbol{%
\tilde{H}}_{N+1}%
\end{array}%
\right),
\end{equation}
where $\boldsymbol{\tilde{H}_{j}}=\boldsymbol{H_{j}}$ for $j=1, 2, ..., N$
in the central region. Proceed by using Eq .~\ref{Greenfun} directly, we can
obtain the desired recursive Green's function ~$G_{S}$~.

\section{RESULTS AND DISCUSSIONS}

\begin{figure*}[tbp]
\includegraphics [width=\columnwidth]{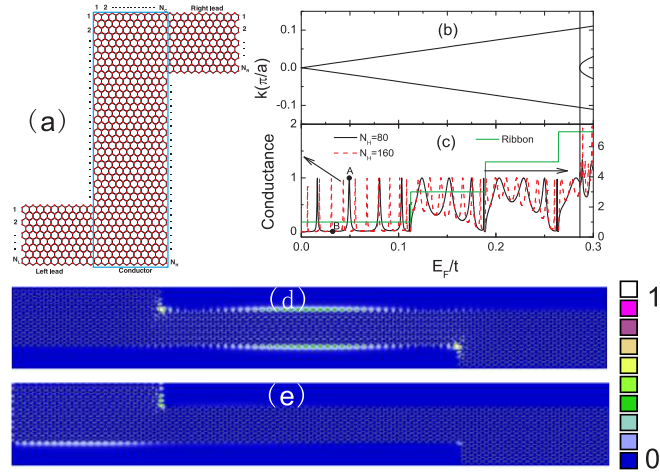}
\caption{(Color online) (a) Schematic diagram of the double-bended graphene
nanoribbon structure. (b) The band structure of the perfect armchair
graphene nanoribbon (AGNR) with the same middle conductor width ($N_{c}$).
(c) The conductance as a function of the Fermi energy through the
double-bended structure, the black-solid and red-dashed lines correspond to
the different conductor lengths $N_{H}$=80 and 160, respectively, the green
line is for the perfect zigzag graphene nanoribbon. (d) and (e) show the
density distributions of quasi-bound electron states marked by the points A
and B in Fig. \protect\ref{Fig1} (c) for $N_{H}=80$ double-bended structure.
$N_{L}$=$N_{R}$=20 and $N_{C}$=8.}
\label{Fig1}
\end{figure*}

First, we discuss the transport property through the double-bended
(S-shaped) GNR structure as shown in Fig. \ref{Fig1} (a) for reference
purposes. Fig. \ref{Fig1} (c) shows the conductance as a function of the
Fermi energy for different lengths of the AGNRs. There are the strong peaks
in the conductance spectrum corresponding to the resonant tunneling process
through the double-bended GNR structure. The underlying physics of such
resonant tunneling process is that there are quasi-bound states in the
middle conductor region which consists of the armchair and zigzag
nanoribbons, the resonant tunneling occurs when the Fermi energy is equal to
the energies of the quasi-bound states. The quasi-bound states arise from
the propagation of electrons back and forth in the armchair region between
the zigzag leads. The conductance of the perfect ZGNR displays a step-like
feature, which corresponds to the opening of the different channels as the
Fermi energy increases (see the green line in Fig. ~\ref{Fig1}(c)). Note
that we did not introduce any external electric field and/or gate to
generate the barrier in the middle part. Therefore, this feature arises from
the geometry effect, i.e., the different edges of the ZGNR leads and the
AGNR conductor. As the Fermi energy increases, more incident channels are
opened, the effect of the edge state becomes weaker, and the ratio between
peaks and valleys of the conductance decreases (see the black and red lines
in Fig. ~\ref{Fig1}(c)). When the Fermi energy increases further and even
becomes higher than the bottom of the second subband of the middle AGNR
conductor (see the black vertical line in Fig. \ref{Fig1} (b)), the
oscillation of the conductance becomes more significant. Fig. \ref{Fig1} (c)
indicates that there are more tunneling peaks as the AGNR length increases.
The incident electron can be completely reflected when the incident energy
is less than the bottom of the second subband of the left ZGNR lead
corresponding to the case of single mode incidence.

\begin{figure}[tbp]
\includegraphics [width=\columnwidth]{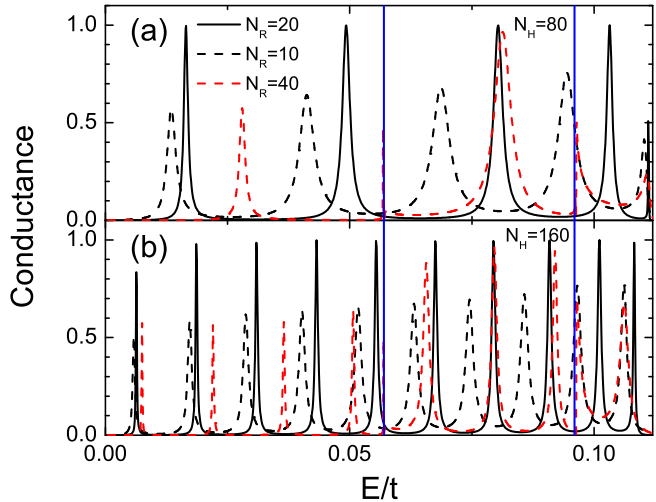}
\caption{(Color online) (a) and (b) The conductance as a function of the
Fermi energy for $N_{H}=80$ and $160$, respectively. The different lines
correspond to the different widths of the right lead for fixed width of the
left lead $N_{L}$=20. The blue vertical lines correspond to the bottom of
the second and third subbands for the perfect ZGNR with the same width of
the right lead $N_{R}$=40.}
\label{Fig2}
\end{figure}

To obtain a clear physical picture, we plot the distribution of electron
states at specific Fermi energies $E_{F}$ corresponding to the points A and
B marked in Fig. \ref{Fig1} (c). Figure \ref{Fig1} shows the density
distribution of the quasi-bound state in the middle AGNR part that
contributes to the resonant tunneling (see the point A in Fig. \ref{Fig1}
(c)). The electrons mainly localize at the edges of the middle AGNR, at the
corner between the AGNR and ZGNR leads, and on the edge of AGNR. This
quasi-bound state is quite different from the electron states of a perfect
AGNR where no edge state appears due to the two different kinds of carbon
atoms at the edges. But for the quasi-bound state, the electron state
strongly localizes at the edges and especially at one kind of carbon atoms
(see Figs. \ref{Fig1} (d) and (e)). This special quasi-bound state in the
middle part contributes dominantly to the resonant tunneling process. We
also plot the density distribution of the quasi-bound state corresponding to
the completely reflection case (see point B in Fig. \ref{Fig1}(c)) in Fig. %
\ref{Fig1}(e). For this case, the electron state localizes at the edge of
the left semi-infinite ZGNR lead near the interface between the lead and the
middle conductor. In this case, the AGNR conductor can be used as an
electron reflector. For the usual graphene nanoribbon, there is no
quasi-bound state in the middle region since no armchair nanoribbon locates
in between the zigzag leads. Therefore we can only observe the step-like
feature of the conductance which corresponds to the opening of the channels
with higher energies. (see the green line in Fig. 3(b))

\begin{figure}[tbp]
\includegraphics [width=\columnwidth]{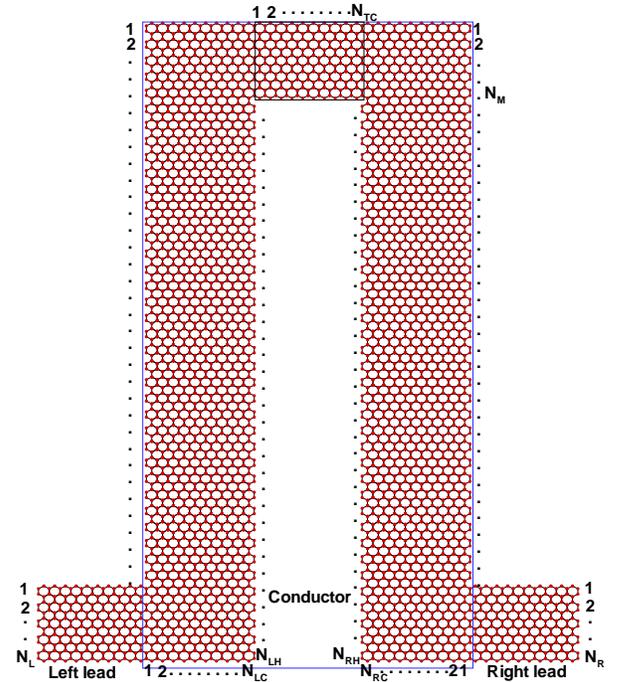}
\caption{(Color online) Schematic diagram of the U-shaped graphene
nanoribbon structures which is composed of two double-bended nanoribbons.
The width of the left (right) ZGNR lead is denoted by $N_{L(R)}$ in the unit
of the number of hexagonal units. The width and length of the left (right)
AGNR are denoted by $N_{LC}$ ($N_{RC}$) and $N_{LH}$ ($N_{RH}$). $N_{M}$ and
$N_{TC}$ correspond to the width and length of the bridge GNR (the black
frame) connecting the left AGNR and the right AGNR. }
\label{Fig3}
\end{figure}

\begin{figure}[tbp]
\includegraphics [bb=0 0 710 980,width=\columnwidth]{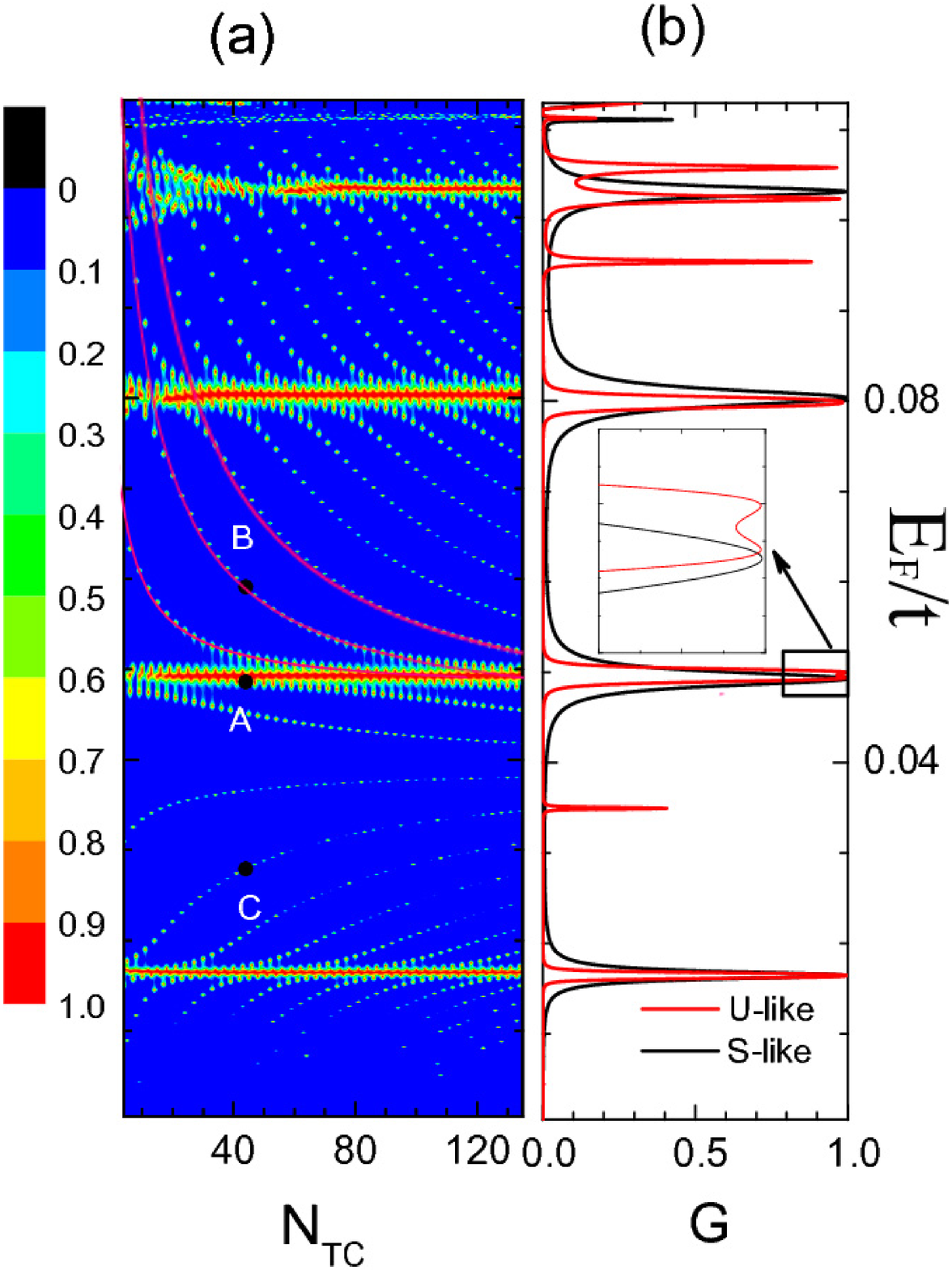}
\caption{(Color online) (a) The contour plot of the conductance as function
of the length of the ZGNR ($N_{TC}$) and Fermi energy ($E_{F}$) for the
U-shaped structure. $N_{L\left( R\right) }=20$, $N_{LC}=N_{RC}=8$, and $%
N_{LH}=N_{RH}=80$. (b) the conductance as a function of the Fermi energy for
the double-blended (S-shaped) GNR structure. $N_{L\left( R\right) }=20$, $%
N_{C}=8$, and $N_{H}=80$. The red line denotes the conductance as a function
of the Fermi energy for the U-shaped structure at $N_{TC}=42$.}
\label{Fig4}
\end{figure}

Next, we show how the width of the right lead affects the resonant tunneling
behavior in Fig. \ref{Fig2}. We fix the width of the left lead at $N_{L}=20$
while tuning the width of the right lead. In Fig. \ref{Fig2}, the black
solid, the black dashed, and the red dashed lines correspond to $N_{R}=20,$ $%
10$, and $40$, respectively. Figure \ref{Fig2} (a) demonstates that the
magnitude of the tunneling peaks decreases whether the width of the right
lead is wider or narrower than that of the left lead. Thus the electrons
cannot be fully transmitted through the nanostructure. The perfect
transmission only takes place when the right lead has the same width as the
left lead. When the width of the right lead is narrower than that of the
left lead (see the black-dashed line in Fig. \ref{Fig2}), the first
tunneling peak is weakened heavily. When the Fermi energy increases, the
magnitude and width of the tunneling peaks increase gradually. These
increases are caused by the enhanced coupling between the quasi-bound states
in the middle part and the leads. Comparing the case with equal widths of
the left and right leads shown in Fig. \ref{Fig2} (a), all tunneling peaks
for the wider (narrower) right leads shift to higher (lower) energy, and
become sparse as the Fermi energy increases. For the wider right lead case,
the conductance shows different features. For the Fermi energy below the
bottom of the second subband in the right lead (see the first vertical green
line), the variation of the magnitude of the tunneling peaks is similar to
the case with the narrower right lead. When the Fermi energy is above the
bottom of the second subband in the right lead, the magnitude of the
tunneling peaks increases abruptly, even approaching to 1, and becomes wider
(see the red dashed peaks in Fig. \ref{Fig4}(b)). Comparing the conductances
through the middle AGNR with different lengths ($N_{H}$) in Fig. \ref{Fig2},
we find that the tunneling peaks become dense as the length $N_{H}$
increases. For longer AGNR ($N_{H}=160$), the variation of the width of the
right lead induces changes in the conductance similar to that of the shorter
AGNR ($N_{H}=80$).

\begin{figure}[tbp]
\includegraphics [width=\columnwidth]{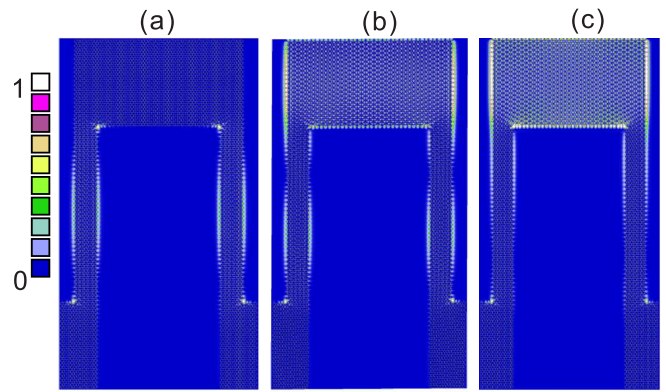}
\caption{(Color online) (a), (b), and (c) show the density distributions of
the quasi-bound electron states corresponding to the points A, B, and C in
Fig. \protect\ref{Fig4} for the $N_{C}=44$ U-type structure. }
\label{Fig5}
\end{figure}

Now we turn to consider the U-shaped structure composed of two double-bend
structures as shown in Fig. \ref{Fig3}. Figure \ref{Fig4} (a) shows the
contour plot of the conductance as a function of the length of the middle
ZGNR\ bridge ($N_{TC}$) and the Fermi energy ($E_{F}$) for the U-shaped
structure. From this figure, the tunneling peaks show a periodic behavior as
a function of the length $N_{TC}$. This period is exactly the same as that
between the metal- and semiconductor-like AGNRs. The resonant tunneling
processes can be divided into two kinds of processes according to the
behavior of tunneling peaks varying with the length $N_{TC}$. The first kind
of tunneling peaks is strong and does not shift when the length of the
middle bridge increases, while the second kind shows the opposite behavior,
i.e., shifting to lower and/or higher energies. Comparing the conductance of
the U-shaped structure with that of one double-bended (S-shaped) structure,
the four peaks in Fig. \ref{Fig4}(b) correspond exactly to the tunneling
peaks of the double-bended structure (S-shaped) (see Figs. \ref{Fig4} (a)
and (b)). Therefore these resonant tunneling processes are caused by the
quasi-bound states localized in the two AGNR regions in the middle conductor
parts, which behave like the resonant tunneling through a single barrier.
This can be understood from the spatial distribution of the quasi-bound
states contributing to the resonant tunneling process marked by point A in
Fig. \ref{Fig4}(a). The quasi-bound electron state localizes at the left and
right AGNR of the middle conductor. From the inset of Fig. \ref{Fig4}(b),
one can see that the first kind resonant peak near E=0.05t split into two
peaks due to the coupling between the two lowest localized states in the
left and right AGNRs through the ZGNR bridge, i.e, the bonding and
antibonding splitting. Therefore the resonant peaks split into two peaks and
show an anticrossing behavior with increasing the length of ZGNR bridge that
describes the coupling strength.\textit{\ }The other resonant tunneling
peaks are shaper and are located between the strong tunneling peaks
mentioned above and vary as the length $N_{TC}$ increases (see Fig. \ref%
{Fig4} (a)). Those peaks is the result of the resonant tunneling processes
through the double barriers, since the corresponding quasi-bound states
localize in the ZGNR bridge of the middle conductor. These quasi-bound
states mainly distribute in the ZGNR bridge, especially at the two vertical
edges of the ZGNR bridge (see Fig. \ref{Fig5} (b) and (c)). The spatial
distributions of the quasi-bound states show the antibonding (Fig. \ref{Fig5}
(b)) and bonding (Fig. \ref{Fig5} (c)) states corresponding to the resonant
tunneling peaks B and C marked in Fig. \ref{Fig5}(a). This bonding and
antibonding feature also explain why the positions of the resonant tunneling
peaks B (C) decrease (increase) as the length $N_{TC}$ increases.

\begin{figure}[tbp]
\includegraphics [bb= 0 0 1360 900,width=\columnwidth]{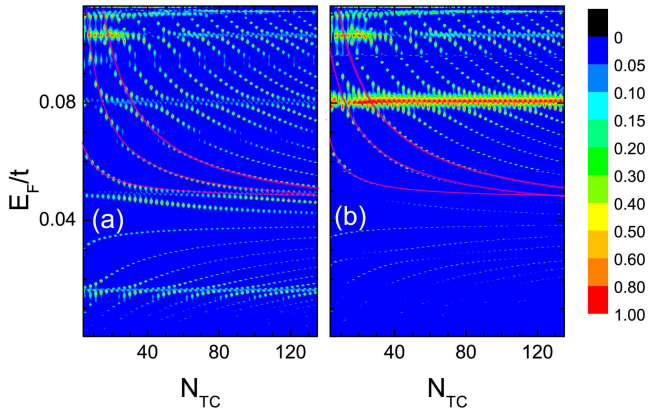}
\caption{(Color online) The same as Fig \protect\ref{Fig4} (a), but (a) and
(b) are for $N_{R}$=10 and 20, respectively.}
\label{Fig6}
\end{figure}

Finally, we investigate the effect of the widths of the right lead on the
transport property. We fix the width of the left lead at $N_{L}=20$ while
tuning the width of the right lead. Figure \ref{Fig6} (a) shows the contour
plot of the conductance as a function of the Fermi energy $E_{F}$ and the
length of the ZGNR bridge $N_{TC}$ for the width of the right lead $N_{R}=10$%
. The figure shows that the first kind of tunneling peaks disappears but the
second kind of tunneling peaks still can be observed, because the latter is
a consequence of tunneling through the quasi-bound states localized in the
middle bridge. The effect of the width of the lead on the second kind of
tunneling processes is much weaker than on the first kind of tunneling
processes. For the wider right lead $N_{R}=40$ in Fig. \ref{Fig6} (b), the
situation is similar with $N_{R}=10$. But when the Fermi energy is larger
than the bottom of the second subband of the right lead, a strong tunneling
process appears at $E_{F}\approx 0.08t$. With a wider or narrower right
lead, the second kind of tunneling processes display similar behavior as the
length ($N_{TC}$) of the bridge increases.\ The tunneling processes induced
by the quasi-bound antibonding states results in tunneling peaks with an
energy that is approximately 0.04t higher. Those peaks shift to lower energy
as the length of the bridge $N_{TC}$ increases. In contrast, opposite
behavior as compared to those induced by the antibonding states, i.e. the
blue shift.

\section{CONCLUSIONS}

In summary, we investigated theoretically the transport properties of
carriers that move through S-and U-shaped graphene nanoribbons. We found
that as function of the Fermi energy the tunneling current can be turned on
and off periodically making this device suitable for transistor action. We
demonstrated that resonant tunneling can be realized utilizing the
geometrical effect of such bended GNR structures even in the absence of any
nanostructured gates. The strong resonant tunneling processes arise from the
quasi-bound states that localize at the middle part and can be clearly seen
in the conductance spectra. The spectra exhibits rich structures coming from
the different kind of quasi-bound states in the middle part. The resonant
tunneling processes can be tuned by changing the Fermi energy and the width
of the left and right leads. Our theoretical results demonstrate that
resonant tunneling can be realized utilizing the geometry of the graphene
nanoribbon without any nanostructured electric gates. We believe that our
theoretical results are interesting both for the basic physics and for the
potential application of electronic devices based on graphene nanoribbons.

\begin{acknowledgments}
This work is supported by the NSF of China Grant No. 60525405 and 10874175,
the Flemish Science Foundation (FWO-Vl) and the Belgian Science policy.
\end{acknowledgments}


\begin{thebibliography}{99}
\bibitem{Novoselov} K. S. Novoselov, A. K. Geim, S. V. Morozov, D. Jiang, Y.
Zhang, S. V. Dubonos, I. V. Grigorieva, and A. A. Firsov, Science \textbf{306%
}, 666 (2004).

\bibitem{QHE} Y. Zhang, Y. Tan, Horst L. Stormer, and P. Kim, Nature \textbf{%
438}, 201 (2005); K. S. Novoselov, Z. Jiang, Y. Zhang, S. V. Morozov, H. L.
Stormer, U. Zeitler, J. C. Maan, G. S. Boebinger, P. Kim, and A. K. Geim,
Science \textbf{315}, 1379 (2007).

\bibitem{Beenakker1} C. W. J. Beenakker, Phys. Rev. Lett. \textbf{97} 067007
(2006).

\bibitem{Sengupta} S. Bhattacharjee and K. Sengupta, Phys. Rev. Lett.
\textbf{97} 217001 (2006).

\bibitem{Katsnelson} M. I. Katsnelson, K. S. Novoselov, and A. K. Geim, Nat.
Phys. \textbf{2}, 620 (2006).

\bibitem{Pereira} J. M. Pereira, P. Vasilopoulos, and F. M. Peeters, Appl.
Phys. Lett. \textbf{90} 132122 (2007).

\bibitem{Egger} A. De Martino, L. Dell'Anna, and R. Egger, Phys. Rev. Lett.
\textbf{98}, 066802 (2007).

\bibitem{Zhai} F. Zhai and Kai Chang, Phys. Rev. B \textbf{77}, 113409
(2008).

\bibitem{Peeters} M. Ramezani Masir, P. Vasilopoulos, A. Matulis, and F.M.
Peeters, Phys. Rev. B \textbf{77}, 235443 (2008)

\bibitem{IBM} Z. H. Chen, Y. M. Lin, M. J. Rooks, and P. Avouris, Physica E
\textbf{40}, 228 (2007).

\bibitem{kim} M. Y. Han, B. Ozyilmaz, Y. B. Zhang, and P. Kim, Phys. Rev.
Lett. \textbf{98}, 206805 (2007).

\bibitem{Nakada} K. Nakada, M. Fujita, G. Dresselhaus, and M. S.
Dresselhaus, Phys. Rev. B \textbf{54}, 17954 (1996).

\bibitem{Wakabayashi} K. Wakabayashi, Phys. Rev. B \textbf{64}, 125428
(2001).

\bibitem{Wakabayashi2} K. Wakabayashi and T. Aoki, Inte. J. Mod. Phys. B
\textbf{16}, 4897 (2002).

\bibitem{Peres} N. M. R. Peres, A. H. Castro Neto, and F. Guinea, Phys. Rev.
B \textbf{73}, 195411 (2006).

\bibitem{Cresti} A. Cresti, G. Grosso, and G. P. Parravicini, Phys. Rev. B
\textbf{77,} 233402 (2008).

\bibitem{Areshkin} D. A. Areshkin, D. Gunlycke, and C. T. White, Nano Lett.
\textbf{7}, 204 (2007).

\bibitem{Yan} Q. M. Yan, B. Huang, J. Yu, F. W. Zheng, J. Zang, J. Wu, B. L.
Gu, F. Liu, and W. H. Duan, Nano Lett. \textbf{7}, 1469 (2007).

\bibitem{Reich} S. Reich, J. Maultzsch, C. Thomsen, and P. Ordej$\acute{o}$%
n, Phys. Rev. B \textbf{66}, 035412 (2002).

\bibitem{butiker} M. B\"{u}ttiker, Y. Imry, R. Landauer, and S. Pinhas,
Phys. Rev. B \textbf{31}, 6207 (1985).

\bibitem{Ando} T. Ando, Phys. Rev. B \textbf{44}, 8017 (1991).

\bibitem{Ozyilmaz} B. Ozyilmaz, P. Jarillo-Herrero, D. Efetov, D. A. Abanin,
L. S. Levitov, and P. Kim, Phys. Rev. Lett. \textbf{99} 166804 (2007).
\end{thebibliography}
\end{document}